\documentclass[12pt]{article}
\usepackage[cp1251]{inputenc}
\usepackage{amstext}
\def\be{\begin{eqnarray}}
\def\ee{\end{eqnarray}}
\def\ba{\begin{array}}
\def\ea{\end{array}}

\begin{document}
\begin{center}
{\Huge\bf Structures of General Relativity in \\ Dilaton-Maxwell Electrodynamics} 
\end{center}

\begin{center}

O.V. Kechkin$^{[1],[2]}$ and P.A. Mosharev$^{[2]}$

\end{center}

\begin{center}
$[1]$ D.V. Skobeltsyn Institute of Nuclear Physics, M. V. Lomonosov Moscow State University, 119234, Moscow, Russia,\\
$[2]$ Faculty of Physics, M. V. Lomonosov Moscow State University, 119991, Moscow, Russia,\\
E-mail: kechkin@srd.sinp.msu.ru,\quad moscharev.pavel@physics.msu.ru

\end{center}

\begin{center}
{\Large\bf Abstract}
\end{center}

It is shown that electro (magneto) static sector of Maxwell's electrodynamics coupled to the dilaton field in a string theory form possesses the symmetry group of the stationary General Relativity in vacuum. Performing the Ernst formalism, we develope a technique for generation of exact solutions in  this modified electrodynamics on the base of the normalized Ehlers symmetry transformation. In the electrostatic case, we construct and study a general class of spherically symmetric  solutions  that describes a point-like sourse of the Coulomb type. It is shown that this source is characterized by asymptotical freedom of the electrostatic interaction at short distances. Also it is established that the total electrostatic energy of this source is finite and inversely proportional to the  dilaton-Maxwell coupling constant.

\begin{center}
\end{center}

\begin{center}
\end{center}

\section*{Introduction}

Dilaton generalization of the standard Maxwell electrodynamics is predicted by a wide class of Grand Unification Theories. Historically the first example of this phenomenon gives a multi-dimensional Kaluza-Klein theory with toroidal compactification of the extra dimensions \cite{KK}. A similar situation occurs in supergravity  and in the superstring theory \cite{SG}--\cite{SS}. Specification of Grand Unification scheme allows one to fix the value of dilaton-Maxwell coupling constant, which turns out to be nontrivial. Dilaton-Maxwell electrodynamics (DME) is significantly non-linear theory. The study and comparison of DME with other non-linear generalizations of Maxwell electrodynamics is an important and interesting problem \cite{NED-1}-\cite{NED-3}. 

In this work we consider DME with negative kinetic term for the  dilaton field. For example, DME of this form appears in the Kaluza-Klein theory  as a result of compactification of extra time dimension. We show that such DME is the closest one to General Relativity (GR) in its formal properties. Namely, we have established a correspondence between electrostatics (and magnetostatics) of DME and stationary General Relativity. This correspondence includes the identification of groups of hidden symmetries of these systems that allows the use of methods of generation of exact solutions of General Relativity in DME. In particular, we have developed the Ernst potential formalism in static DME, and built a technique of generating asymptotically trivial solutions in DME using its charging symmetries (in the sense of Kinnersley \cite{Kinnersley-1}--\cite{Kinnersley-4}). In fact, we have performed normalized Ehlers transformation to generate a class of solutions invariant under the action of the charging symmetry subgroup of the  system \cite{Ehlers}, \cite{OK-0}. We take the harmonic dilaton field and the electromagnetic zero background as a basis for generating solutions of DME.

Applying the generation procedure we have constructed a family of spherically symmetric solutions of DME, describing the modified source of the Coulomb type in this theory. This family forms invariant class of solutions of the group of charging symmetries, it corresponds to the Taub-NUT solution in General Relativity \cite{Taub-NUT-1}-\cite{Taub-NUT-2}. It is shown that a more interesting case of electrostatic source is characterized by an electric field, which is finite everywhere and has Coulomb behavior at spatial infinity. Moreover, the electric field of this source disappears in the center of the distribution, i.e. the corresponding interaction possesses the property of asymptotical freedom at short distances. This behavior is characteristic of theories with non-Abelian gauge symmetry \cite{YM}; thus, DME with an Abelian gauge structure is similar to them because of the appearance of hidden non-Abelian symmetry at the static level. 

Then, the total energy of this source contains a divergent part, which does not depend on the interaction between the fields of the theory (and has not, apparently, a significant physical meaning). After regularization it turns out that the finite mass-energy of the system is inversely proportional to the value of dilaton-Maxwell coupling constant. It is shown that the standard electrostatic energy of this source has no divergences, and  possesses the same non-perturbative dependence on the interaction constant  between the fields of the theory. Thus, the generalized Coulomb source presented in this paper demonstrates soliton properties as relevant analogies from the theories with non-Abelian gauge symmetry group \cite{Soliton}.

\section{DME in 4 and 3 Dimensions}

Dilaton-Maxwell electrodynamics is the four-dimensional theory on a flat background with the Lagrangian
\be\label{DME-1-1}
L=-\frac{1}{4}\,e^{-2\alpha\phi}F_{\mu\nu}F^{\mu\nu}-2\partial_{\mu}\phi\,\partial^{\mu}\phi,\ee
where $F_{\mu\nu}=\partial_{\mu}A_{\nu}-\partial_{\nu}A_{\mu}$ denotes the electromagnetic tensor, whereas $A_{\mu}=A_{\mu}\left(x^{\lambda}\right)$ and $\phi=\phi\left(x^{\lambda}\right)$ are the electromagnetic potential and dilaton field, respectively. Here $\alpha ={\rm const}$ is a dilaton-Maxwell coupling, and Greek indices $\mu,\nu,\lambda=0,...,3$ are raised and lowered using the Minkowski metric $\eta^{\mu\nu}=\eta_{\mu\nu}={\rm diag}\left(1,\,-1,\,-1,\,-1\right)$. We consider DME with negative kinetic term for the dilaton: as it is shown below, this form of DME is the closest one to the General Relativity. It is important to note that the group of DME symmetries in four dimensions includes the dilaton shift
$\phi\rightarrow\phi+\Lambda,\,\, A_{\mu}\rightarrow e^{\alpha\Lambda}A_{\mu}$ (where $\Lambda={\rm const}$) in addition to the `standard'\, Poincare and Abelian gauge symmetry transformations.

The system of Euler-Lagrange equations, which corresponds to Eq. (\ref{DME-1-1}), reads: 
\be\label{DME-1-3}
&&\partial_{\mu}\left(e^{-2\alpha\phi}F^{\mu\nu}\right)=0,\\
\label{DME-1-4}
&&\partial_{\mu}\partial^{\mu}\phi+\frac{\alpha}{8}e^{-2\alpha\phi}F_{\mu\nu}F^{\mu\nu}=0.
\ee
Considering the stationary case, where $\partial_0 A_{\mu}=\partial_0\phi=0$, i.e. $A_{\mu}=A_{\mu}\left(x^k\right)$ and  $\phi=\phi\left(x^k\right)$ (with  $k=1,2,3$),  one obtains from Eq. (\ref{DME-1-3}) for $\nu=0$:
\be\label{DME-1-5}\nabla\left(e^{-2\alpha\phi}\nabla A^0\right)=0,\ee
where $\nabla=\left\{\partial_k\right\}$. Then, the sector $\nu=k$ of  Eq. (\ref{DME-1-3}) becomes
\be\label{DME-1-6}\nabla\times\left(e^{-2\alpha\phi}\nabla\times\vec A\right)=0\ee
in the stationary case, where $\vec A=\left\{A^k\right\}$,  while Eq. (\ref{DME-1-4}) takes the following form: 
\be\label{DME-1-7}\nabla^2\phi+\frac{\alpha}{4}e^{-2\alpha\phi}\left[
\left(\nabla A^0\right)^2-\left(\nabla\times\vec A\right)^2\right]=0.\ee
By a direct calculation one can verify that Eqs. (\ref{DME-1-5})--(\ref{DME-1-7}) are the Euler-Lagrange equations for an effective three-dimensional system with the Lagrangian 
\be\label{DME-1-8}L=2\left(\nabla\phi\right)^2+\frac{1}{2}e^{-2\alpha\phi}
\left[
\left(\nabla A^0\right)^2-\left(\nabla\times\vec A\right)^2\right].\ee

Alternative description of this system is based on the use of magnetic potential $\tilde A^0$, which is introduced by the relation 
\be\label{DME-1-9}\nabla\tilde A^0=-e^{-2\alpha\phi}\nabla\times\vec A,\ee
which is consistent in view of Eq. (\ref{DME-1-6}). From Eq. (\ref{DME-1-9}) we obtain the relation 
\be\label{DME-1-10}\nabla\left(e^{2\alpha\phi}\nabla\tilde A^0\right)=0,\ee
whereas from Eq. (\ref{DME-1-7}) (and taking into account Eq. (\ref{DME-1-9})) we come to the equation 
\be\label{DME-1-11}\nabla^2\phi+\frac{\alpha}{4}\left[e^{-2\alpha\phi}
\left(\nabla A^0\right)^2-e^{2\alpha\phi}\left(\nabla\tilde A^0\right)^2\right]=0.\ee
Finally, it is easy to check that Eqs. (\ref{DME-1-5}), (\ref{DME-1-10}) and (\ref{DME-1-11}) correspond to the effective Lagrangian
\be\label{DME-1-12}\tilde L=2\left(\nabla\phi\right)^2+\frac{1}{2}\left[e^{-2\alpha\phi}
\left(\nabla A^0\right)^2+e^{2\alpha\phi}\left(\nabla\tilde A^0\right)^2\right].\ee
From Eq. (\ref{DME-1-12}) it is seen that the stationary DME is a three-dimensional $\sigma$-model on a flat background in terms of the potentials $\phi,\, A^0$ and $\tilde A^0$. This $\sigma$-model is nonlinear one if $\alpha\neq 0$. 

\section{Static DME / stationary GR duality}
Let us now consider  separately the electric (e) and magnetic (m) sectors of the stationary DME with $\alpha\neq 0$. Thus, from this point on we deal with the eletcro- and/or magnetostatics defined by the restrictions
\be\label{DME-1-13}&{\rm e}&:\,\,\, \tilde A^0=0, \qquad A^0= A\neq 0,\\
\label{DME-1-13-2}&{\rm m}&:\,\,\, A^0=0, \qquad \tilde A^0= A\neq 0\ee
for the e/m special cases, respectively. It is easy to verify that the subsystems (\ref{DME-1-13}) and (\ref{DME-1-13-2}) free of any additional constraints (Eqs. (\ref{DME-1-10})/(\ref{DME-1-5}) are identities on Eqs.  (\ref{DME-1-13})/(\ref{DME-1-13-2})). Using Eq. (\ref{DME-1-12}), it is not difficult to prove that the corresponding Lagrangian has the following form: 
\be\label{DME-1-14} \tilde L_{\rm e/m}=\frac{4}{\alpha^2}\,\tilde L_{\rm GR},\ee
where
\be\label{DME-1-15}\tilde L_{\rm GR}=\frac{1}{2}\,f^{-2}\left[\left(\nabla f\right)^2+\left(\nabla\chi\right)^2\right],\ee
and
\be\label{DME-1-16}f=e^{\pm\alpha\phi},\qquad \chi =\frac{\alpha}{2}A.\ee
The statement is that the same Lagrangian describes General Relativity (GR) in vacuum in the stationary case, with a corresponding interpretation of the potentials $f$ and $\chi$ \cite{Ernst-1}--\cite{GR-3}. Namely, in this theory the metric $g_{\mu\nu}=g_{\mu\nu}\left(x^{\lambda}\right)$ is Ricci-flat (i.e., $R_{\mu\nu}=0$, where $R_{\mu\nu}$ is the Ricci tensor).  Parameterizing this four-dimensional metric in the form $ds^2=g_{\mu\nu}dx^{\mu}dx^{\nu}=
f\left(dx^0+\omega_kdx^k\right)^2-f^{-1}ds_3^2$, where $ds_3^2=h_{kl}dx^kdx^l$ is the three--dimensional line element, let us consider the stationary case with $\partial_0f=\partial_0\omega_k=\partial_0h_{kl}=0$. Then, introducing the rotational potential $\chi$ according to the relation $\nabla\chi=-f^2\nabla\times\vec{\omega}$, one leads to the effective three-dimensional theory with the action \be\label{DME-1-20}\tilde S=\int ds_3^2\,h^{1/2}\left(-R_3+\tilde 
L_{\rm GR}\right),\ee where $h={\rm det}\,h_{kl}$, $R_3$ is the curvature scalar constructed over the 3-metric $h_{kl}$, and the Lagrangian $\tilde L_{\rm GR}$ is given by Eq. (\ref{DME-1-15}).  

It is interesting to note that the effective three-dimensional action of General Relativity written in terms of the original metric, i.e., \be\label{DME-1-21}S=\int d^3x\,h^{1/2}\left(-R_3+ L_{\rm GR}\right),\ee where \be\label{DME-1-22}L_{\rm GR}=\frac{1}{2}\,\left[f^{-2}\left(\nabla f\right)^2-f^2\left(\nabla\times\vec{\omega}\right)^2\right],\ee and $\vec{\omega}=\left\{\omega^k\right\}$, exactly corresponds to the DME magnetostatics in view of indentifications $\vec\omega=\alpha/2\, \vec A$ and $L_{\rm GR}=\alpha^2/4\,\, L_m$.

\section{Generation of DME solutions using GR hidden symmetries}

Static dilaton-Maxwell electrodynamics possesses a group of hidden symmetries of the stationary General Relativity, as it follows from results of the previous section. Here we mean symmetries of the point type (in fact, the isometries of target space of the $\sigma$-model (\ref{DME-1-15})--(\ref{DME-1-16})) which mix potentials $\phi$, $A$ in  DME / $f$, $\chi$ in GR and do not affect the three-dimensional coordinates $x^k$. Moreover, we are interested in its charging symmetry subgruop in view of possible applications to generation of classes of exact solutions trivial at spatial infinity \cite{Kinnersley-4} (see Appendix for details). 

The group of charging symmetries of static DME  is one-parametric. Its action on the potentials of the theory is given by the map \be\label{DME-1-32}\phi&\rightarrow& \phi\mp \frac{1}{\alpha}\log \left[\left(\cos\frac{\lambda}{2}+\frac{\alpha}{2}\,A\,\sin \frac{\lambda}{2}\right)^2+e^{\pm2\alpha\phi}\sin^2\frac{\lambda}{2}\right],
\nonumber\\
A&\rightarrow& \frac{A\cos\lambda-\frac{1}{\alpha}\sin\lambda\left(1-e^{\pm2\alpha\phi}-
\frac{\alpha^2}{4}A^2\right)}{\left(\cos\frac{\lambda}{2}+\frac{\alpha}{2}\,A\,\sin \frac{\lambda}{2}\right)^2+e^{\pm2\alpha\phi}\sin^2\frac{\lambda}{2}},\ee as it follows from Eqs. (\ref{DME-1-16}), (\ref{DME-1-25}), (\ref{DME-1-30}) and (\ref{DME-1-31}). It is not difficult to prove that this map actually preserves the vacuum point $\phi=A=0$.
Then, considering monopole terms in multipole expansion of the fields near to the spatial infinity $r\rightarrow  +\infty$, \be\label{DME-1-33} \phi=\frac{Q_d}{r}+\dots,\qquad A=\frac{Q_{e/m}}{r}+
\dots,\ee and using Eq. (\ref{DME-1-32}), one leads to the following transformation relations: 
\be\label{DME-1-34}Q_d\,\,\,\,\,\,&\rightarrow& Q_d\,\,\,\,\,\cos\lambda\mp \frac{1}{2}Q_{e/m}\sin\lambda, \nonumber\\  Q_{e/m}&\rightarrow& Q_{e/m}\cos\lambda\pm 2\,\,Q_d\,\,\,\,\,\sin\lambda\ee
for the dilaton $Q_d$ and electric/magnetic $Q_{e/m}$ charges. It is seen that the group of charging symmetries is realized as the group of rotations $SO(2)$ on the charge coordinates $2Q_d$ and $Q_{e/m}$; $\lambda$ is a natural parameter of this group here. Introducing the linearizing potential $z$ according to Appendix, one obtains the expansion $z=\mp {\alpha Q}/{4r}+\dots$  for the fields (\ref{DME-1-33}), where $Q=2Q_d\pm iQ_{e/m}$. It is easy to check that the map (\ref{DME-1-34}) takes explicit $U(1)$ form $Q\rightarrow e^{i\lambda}Q$ in terms of the complex charge $Q$. Then, it is important to note that the charge quantity 
$QQ^*=Q_{e/m}^2+4Q_d^2$ 
is invariant under the chargithng symmetry transformation.
  
Let us now formulate generation procedure based on the charging symmetry (\ref{DME-1-32}). It is clear, that the static DME allows the ansatz with $\phi=\Upsilon$ and $A=0$, where the potential $\Upsilon$ is harmonic, 
\be\label{DME-1-40}\Delta\Upsilon=0,\ee
as it follows from the equations of motion corresponding to the Lagrangian (\ref{DME-1-12}) under the restriction (\ref{DME-1-13}). This ansatz can be used as the base for generation; the generated fields will be asymptotically trivial if the harmonic function $\Upsilon$ possesses the same property. Then from Eq. (\ref{DME-1-32}) we get: \be\label{DME-1-41}\phi=\mp\frac{1}{\alpha}\,\log\Pi,\qquad
A=\pm\frac{2\sin\lambda}{\alpha}\,\cdot\frac{\sinh\alpha\Upsilon}{\Pi},\ee where \be\label{DME-1-42}\Pi=\cosh\alpha\Upsilon\mp\cos\lambda\cdot\sinh\alpha\Upsilon.\ee One can check that the generated potential $A$ is enclosed between its asymptotic values $\mp 2\alpha^{-1}\cot\left(\lambda/2\right)$ and $\pm 2\alpha^{-1}\tan\left(\lambda/2\right)$ as it follows from the simple algebraic analysis. Then, for the electric $\vec E=-\nabla A^0$ and magnetic  $\vec H=\nabla\times \vec A$ fields of the generated class of solutions (\ref{DME-1-41}) one obtains the following expressions:
\be\label{DME-1-43}\vec E=-2\sin\lambda\,\frac{\nabla\Upsilon}{\,\,\,\Pi^2},\qquad \vec H=2\sin\lambda\,\nabla\Upsilon.\ee

\section{General DME central sourse}

Let us choose Coulomb solution $\Upsilon={Q_0}/{r}$, where $Q_0={\rm const}$, as a concrete generation base in Eqs. (\ref{DME-1-41})--(\ref{DME-1-43}). Then one obtains:
\be\label{DME-1-44-1}\vec E=\Pi^{-2}\,\cdot \frac{Q_e\vec r}{r^3},
\qquad \vec H=\frac{Q_m\vec r}{r^3}, \quad e^{-\alpha\phi}=\Pi,\ee
where 
\be\label{DME-1-44-2}\Pi=\cosh\left(\frac{\alpha\sqrt{Q_e^2+4Q_d^2}}{2r}\right)-\frac{2Q_d}{\sqrt{Q_e^2+4Q_d^2}}\sinh\left(\frac{\alpha\sqrt{Q_e^2+4Q_d^2}}{2r}\right)\ee
Here the charges for the generated class of solutions are equal to 
$Q_d=Q_0\cos\lambda$ and $Q_{e/m}=\pm 2Q_0\sin\lambda$,
as it follows from Eq. (\ref{DME-1-34}). Then, considering more interesting electrostatic sector of the theory, one concludes that the electric field $\vec E$ has a standard Coulomb behavior at the spatial innity, and it vanishes at the center of its distribution (i.e., $\vec E\left(r\right)\rightarrow 0$ at $r\rightarrow 0$).
Moreover, it is not difficult to prove that this modified Coulomb source has a finite value of the  electric field everywhere. 

Let us calculate now the total energy ${\cal E}_{tot}=\int d^3x\,\left[-2\left(\nabla\phi\right)^2+1/2\,e^{-2\alpha\phi}\vec E^2\right]$ for the modified Coulomb solution in the electrostatic case. The result has the following form: ${\cal E}_{\rm tot}=\bar {\cal E}_{\rm tot}+\tilde {\cal E}_{\rm tot}$, where the contribution of $\bar {\cal E}_{\rm tot}=-2\int d^3x\,\left(\nabla\Upsilon\right)^2=-2\pi\left(Q_e^2+4Q_d^2\right)\int_0^{+\infty}dr/r^2$ diverges at small distances, whereas the term \be\label{DME-1-49-2}\tilde {\cal E}_{\rm tot}=4\sin^2\lambda\int d^3x\,\frac{\left(\nabla\Upsilon\right)^2}{\Pi^2}=\frac{8\pi Q_e^2}{\alpha\left(
\sqrt{Q_e^2+4Q_d^2}-2Q_d\right)}\ee is finite. It is interesting to note that the term $\bar {\cal E}_{\rm tot}$ is an invariant of the group of charging symmetries, it describes some `irreducible part'\, of the total energy of the system. It can be assumed that the contribution of $\bar {\cal E}_{\rm tot}$ to the total energy should not affect the physical consequences of the theory, and that it can be resolved after the corresponding regularization. In this case, the term 
$\tilde {\cal E}_{\rm tot}$ will describe the full physical energy of the system. 

Finally, let us consider the dynamics of a point particle with charge $q_e$ and mass $m$, interacting with the fields of DME through the conventional Lorentz force. The corresponding equation of motion has the standard form $dp^{\mu}/ds=q_eF^{\mu\nu}u_{\nu}$, where $p^{\mu}=mu^{\mu}$ and  $u^{\mu}$ is the 4-velocity of a particle. Using such particles as a test, it is possible to  determine the distribution of electric field $\vec E$ (but not information about a scalar field $\phi$, which acts in this dynamics as the `phantom'). One can then calculate the standard energy of the electric field 
${\cal E}_{\rm e}=1/2\int d^3x \vec E^2$ of the system; the result reads:
\be\label{DME-1-50}{\cal E}_{\rm e}=2\sin^2\lambda\int d^3x \frac{\left(\nabla\Upsilon\right)^2}{\Pi^4}=
\frac{8\pi}{3\alpha}\frac{\sqrt{Q_e^2+4Q_d^2}+2Q_d}{\sqrt{Q_e^2+4Q_d^2}-2Q_d}
\left(\sqrt{Q_e^2+4Q_d^2}-Q_d\right).\ee
Note that ${\cal E}_{\rm e}\sim 1/\alpha$, i.e. the results of calculation of the energies (\ref{DME-1-49-2}) and (\ref{DME-1-50}) are nonperturbative with respect to dilaton-Maxwell coupling constant $\alpha$. In the special case of $Q_d=0$, which is the closest to the standard Coulomb source, one obtains: $\tilde {\cal E}_{\rm tot}=8\pi\left|Q_e\right|/\alpha$ and
${\cal E}_{\rm e}=8\pi\left|Q_e\right|/3\alpha$. Thus, only the third part from the total energy of the source can be interpreted by this observer as an electric `visible'\, energy, whereas the remaining two-thirds of the total mass-energy appear as dark matter \cite{DM-1}.

\section*{Appendix: Stationary GR and its hidden symmetries}

Ernst representation is particularly convenient in the study of hidden symmetries of the stationary General Relativity. This representation is based on the use of the complex Ernst potential \cite{Ernst-1}--\cite{Ernst-2}
\be\label{DME-1-25} E=f+i\chi,\ee in terms of which Eq. (\ref{DME-1-15}) takes the following form: 
\be\label{DME-1-26}\tilde L_{\rm GR}=2\frac{\nabla\! E\cdot\!\nabla\! E^*}
{\left(E+E^*\right)^2}.\ee
This Lagrangian  possesses the so-called shift and scale transparent symmetry transformations: \be\label{DME-1-27}E\rightarrow E+i\lambda_1,\qquad E\rightarrow \lambda_2E,\ee where $\lambda_1$ and $\lambda_2$ are real constants ($\lambda_2>0$). Then, the inversion $E\rightarrow E^{-1}$ is a discrete symmetry of Eq. (\ref{DME-1-26}). This inversion generates the Ehlers transformation from the shift symmetry:
\be\label{DME-1-29}E^{-1}\rightarrow E^{-1}+i\lambda_3,\ee
where $\lambda_3={\rm const}$ \cite{Ehlers}. Eqs. (\ref{DME-1-27}) and (\ref{DME-1-29}) give the total group of hidden symmetries of the stationary General Relativity which do not affect the coordinates of the physical space-time.  

Then, the class of asymptotically flat solutions, with $f=1,\,\chi=0$ (i.e., with $E=1$)
at the spatial infinity, is of special interest. It is easy to check that the Ehlers transformation with an arbitrary value of the parameter $\lambda_3=\lambda_{\rm _{Eh}}$, followed by the scale transformation performed with $\lambda_2=1+\lambda_{\rm _{Eh}}^2$ and the shift one with $\lambda_1=\lambda_{\rm _{Eh}}$, preserve the property of asymptotic flatness.
This sequence of transformations leads to the map \be\label{DME-1-30}E\rightarrow \frac{E+i\lambda_{\rm _{Eh}}}{1+i\lambda_{\rm _{Eh}} E},\ee which is called a normalized Ehlers symmetry transformation. Eq. (\ref{DME-1-30})  describes the charging symmetry map of the general form in stationary General Relativity. Introducing new group parameter 
\be\label{DME-1-31}\lambda_{\rm _{Eh}}=-\tan\left({\lambda}/{2}\right)\ee
and the complex potential $z=\left(1-E\right)/\left(1+E\right)$, one gets the map
$z\rightarrow e^{i\lambda}z$ from Eq. (\ref{DME-1-30}) \cite{Mazur}. It is seen that $z$ is a linearizing potential for the group of charging symmetries of the system, whereas $\lambda$ is a natural parameter for this group. 

Kaluza-Klein theory, supergravity and superstring theory give a wide class of systems admitting the $\sigma$-model representation. Some of these systems admit the Ernst potential formulation, which is very convenient for the analysis of hidden symmetries \cite{OK-1-0}--\cite{OK-1-2}.  In \cite{OK-1-1}--\cite{OK-1-1-1} it was found generalized Ehlers symmetry for the corresponding theories. Then, charging symmetry transformations and potential representations for these systems were established in \cite{OK-2-1}--\cite{OK-2-2}. The formalism developed was used to generate charging symmetry invarant classes of asymptotically trivial (asymptotically flat) solutions in these nontrivial modifications of General Relativity \cite{OK-3-1}--\cite{OK-3-2}.

\section*{Conclusion}

Thus, there is an interesting duality between dilaton modification of Maxwell's electrodynamics and Einstein's General Relativity. For its implementation it is important that the dilaton field has provided a negative contribution to the total energy of the system just as in the case of the Einstein's gravity (in the approach with the energy-momentum pseudotensor for the gravitational field). This duality allows one to use the generation procedures of stationary General Relativity to construct exact solutions in a static DME. However, such a construction has no the character of an immediate interpretation of the results of General Relativity in DME terms. The reason for this phenomenon is that the effective three-dimensional DME lives on a flat background, while it is curved in the GR theory case (and, moreover, this background is dynamic one in General Relativity). Nevertheless, these systems have the isomorphic $\sigma$-model target spaces and, therefore, they have exactly identical groups of isometries for them. This fact allowed one to apply non-trivial symmetries of General Relativity and to obtain new solutions in dilaton-Maxwell's electrodynamics. It is important to note that the explicit form of the GR and DME solutions, corresponding to each other, turns out to be completely different. For example, a modified Coulomb's solution (\ref{DME-1-44-1})--(\ref{DME-1-44-2}) is not similar to the class of Taub-NUT solutions despite the fact that these solutions exactly correspond to each other under duality (\ref{DME-1-16}). The reason, as we have already noted, lies in a radically different three-dimensional metric backgrounds in these two theories.

It is demonstrated that the presence of dilaton in theory leads to non-trivial results, such as the appearance of asymptotical freedom for the interaction at small distances and to soliton-like properties of the source of this interaction. Thus, a relatively simple theory with gauge group $U(1)$ demonstrates the properties of non-Abelian models with the Yang-Mills gauge fields. In addition, it is shown that theories with dilaton  provide a natural possibility of interpreting some results in terms of dark matter.

\section*{Acknowledgments} We thank our colleagues for encouraging.

\end{document}